\RequirePackage{fix-cm}
\documentclass[smallextended]{svjour3}       
\smartqed  
\usepackage{graphicx}
\usepackage{multirow}
\usepackage{epsfig}
\journalname{}

\title{Parity proofs of the Kochen-Specker theorem based on the 120-cell}

\author{Mordecai Waegell and P.K. Aravind }

\authorrunning{M.Waegell, P.K. Aravind}

\institute{M.Waegell, P.K. Aravind \at
Physics Department, Worcester Polytechnic Institute, Worcester, MA 01609, U.S.A.\\
\email{caiw@wpi.edu, paravind@wpi.edu}}

\date{\today}

\begin{document}
\maketitle
\begin{abstract}
It is shown how the 300 rays associated with the antipodal pairs of vertices of a 120-cell (a four-dimensional regular polytope) can be used to give numerous ``parity proofs" of the Kochen-Specker theorem ruling out the existence of noncontextual hidden variables theories. The symmetries of the 120-cell are exploited to give a simple construction of its Kochen-Specker diagram, which is exhibited in the form of a ``basis table" showing all the orthogonalities between its rays. The basis table consists of 675 bases (a basis being a set of four mutually orthogonal rays), but all the bases can be written down from the few listed in this paper using some simple rules. The basis table is shown to contain a wide variety of parity proofs, ranging from 19 bases (or contexts) at the low end to 41 bases at the high end. Some explicit examples of these proofs are given, and their implications are discussed.

\end{abstract}

\section{\label{sec:Intro}Introduction}

In two recent papers \cite{Waegell2011a,Waegell2011b} we showed how two of the exceptional four-dimensional regular polytopes, the 24-cell and the 600-cell, can be used to give a large number of ``parity proofs" of the Kochen-Specker (KS) theorem \cite{KS1967} ruling out the existence of noncontextual hidden variables theories. In this paper we show how the third and most complex of these polytopes, the 120-cell, yields still further proofs of the same kind. Thus our three papers collectively show how these beautiful geometric objects, which have been known since the middle of the 19th century, can be enlisted, if a bit quixotically, in defending the quantum theory against an attack mounted on it by a personage no less than Einstein.\\

The parity proofs based on the 24-cell have their origin in the proofs of the KS theorem given by Peres \cite{Peres1991} and Mermin \cite{Mermin1993}. Mermin's proof was based on sets of commuting observables for a pair of qubits, while Peres' proof was based on a set of 24 states derived from these observables. Kernaghan \cite{Kernaghan1994} later showed that Peres' states contain subsets of 20 that give parity proofs, and Cabello et al \cite{Cabello1996} showed that there are subsets of 18 that do likewise. One of us \cite{Aravind2000} pointed out that the 24-cell, together with its dual\footnote{The dual of a 24-cell is another 24-cell rotated relative to the first (about their common center), with the vertices of the dual being along the same directions as the cell centers of the original, and vice-versa.}, is the natural geometric framework for the system of rays introduced by Peres. The interest of this observation is that it permits a simple geometrical construction to be given \cite{Waegell2011a} of all the $2^{9} = 512$ parity proofs in this system. An exhaustive study of all the KS sets of vectors in the 24-ray Peres set, whether they gave rise to parity proofs or not, has been carried out by Pavi\v ci\'c and his collaborators \cite{Pavicic2010}.\\

The fact that the 24-cell, together with its dual, led to parity proofs suggested that the other four-dimensional regular polytopes might do likewise. The three simpler polytopes (the simplex, the cross polytope and the measure polytope) are too meager to lead to anything, but in \cite{Waegell2011b} we found, to our great surprise, that the 600-cell has a staggeringly large number of parity proofs in it. It should be stressed that while the 600-cell does have twenty five 24-cells in it, none of them is accompanied by its dual, and so there is no overlap between its parity proofs and those of the Peres set. The contrast between the parity proofs in these two systems is very striking: whereas the Peres rays have only six distinct (i.e. unitarily inequivalent) types of parity proofs (and a total of 512 proofs when all their replicas under symmetry are taken into account), the 600-cell has over a hundred distinct types of proofs (and over a hundred million when all replicas under symmetry are taken into account).\\

We were naturally led to ask whether the 120-cell, the most complex of these polytopes, might have any parity proofs in it. The 120-cell is remarkable in having copies of all the smaller polytopes in it. In particular, it has 10 600-cells and 225 24-cells in it. However none of the 24-cells is accompanied by its dual, and so none of the parity proofs of the Peres set is contained in the 120-cell. But all the parity proofs of the 600-cell are contained in the 120-cell (in 10 different incarnations, in fact). The question, then, is whether the 120-cell has any \textit{new} parity proofs in it, i.e., ones that span two or more of its 600-cells. It was far from obvious to us that it should have any proofs of this kind. However we have discovered that it does, and it is the purpose of this paper to report that discovery.\\

The parity proofs provided by the four-dimensional regular polytopes all involve rays in a real four-dimensional space (which is, in fact, the simplest setting in which parity proofs can arise). Let us recall the other types of spaces in which parity proofs have been found. Kernaghan and Peres \cite{KP1995} found a 36-ray 11-basis proof in a real 8-dimensional space which, together with the proofs in the Peres set \cite{Kernaghan1994,Cabello1996}, were the only parity proofs known for many years. Then, a few years back, there was an explosion in our knowledge. It was shown \cite{Waegell2011c,Waegell2012a,Waegell2013} that there exist parity proofs in every complex space $C^{d}$ of dimension $d=2^{N}$ (for $N \geq 2$) that can be derived from suitable subsets of observables of the $N$-qubit Pauli group. This showed that parity proofs are not singular phenomena but occur systematically in the state space of any number of qubits, with the variety and quantity of such proofs increasing sharply with the number of qubits. Very recently, a completely unexpected discovery was made: Lisonek et al \cite{Lisonek2013} found a 21-ray 7-basis proof in a complex 6-dimensional space that is remarkable because it involves the smallest number of bases (seven) known for a parity proof in any dimension and also because it is the first parity proof to be discovered in a dimension not of the form $2^{N}$. This discovery seems to hint at the fact that there may still be things about parity proofs that we do not know.\\

The parity proofs of this paper, like the others that have preceded them, are of interest for a variety of reasons: they can be used to derive state-independent inequalities for ruling out noncontextuality \cite{Cabello2008,Kirchmair} and Bell inequalities for identifying fully nonlocal correlations \cite{Aolita}; they have applications to quantum games \cite{Ambrosio}, quantum zero-error communication \cite{Cubitt}, quantum error correction \cite{Error} and the design of relational databases \cite{Abramsky2012}; and they can be used to witness the dimension of quantum systems \cite{Guhne}.\\

The plan of this paper is as follows. In Sec.\ref{sec:2} we give a simple construction of the rays and bases of the 120-cell based on its symmetries. In Sec.\ref{sec:3} we review the notion of a parity proof and identify substructures within the 120-cell that are more easily searched for such proofs. We then list the various types of proofs we have found, in terms of their symbols (defined below), and give explicit examples of a few of the proofs. Finally, in Sec.\ref{sec:4}, we make some concluding remarks.

\section{\label{sec:2} Geometry of the 120-cell: rays and bases}

The 120-cell \cite{Coxeter} has 600 vertices distributed symmetrically on the surface of a sphere in four-dimensional Euclidean space. The vertices come in antipodal pairs, and the lines through antipodal pairs of vertices define the 300 rays of the 120-cell. We will term any set of four mutually orthogonal rays (or directions) a basis. The 300 rays form 675 bases, with each ray occurring in 9 bases and being orthogonal to its 27 distinct companions in these bases and to no other rays. We will use the symbol $300_{9}$-$675_{4}$ to denote this system of rays and bases, with the left half of the symbol indicating the number of rays (with their multiplicities\footnote{The multiplicity of a ray is the number of bases it occurs in.} as subscripts) and the right half the number of bases (with the number of rays in each basis as a subscript). We will use a similar notation for the other ray-bases systems that will be encountered below. For example, $60_{2}180_{6}$-$300_{4}$ denotes a system of 240 rays and 300 bases, with 60 rays of multiplicity 2 and 180 rays of multiplicity 6. We will only deal with bases of four rays in this paper, so the subscript in the right half of the symbol will always be 4 (and will sometimes be dropped, for brevity).\\

The 120-cell has the property that all the orthogonalities between its rays are represented among its bases. Thus its basis table (i.e., the list of all its bases) contains the same information as its Kochen-Specker diagram\footnote{The Kochen-Specker diagram of a set of rays is a graph whose vertices are the rays and whose edges connect vertices corresponding to orthogonal rays.}. The basis table of the 120-cell is an object of great interest, because it is the structure within which all its parity proofs are embedded. In fact, any parity proof is just some subset of these bases, as we will see in Sec.3.\\

A listing of the full basis table of the 120-cell would take up too much space and is also unnecessary. We will explain how all the bases can be built up by applying suitable symmetry operations of the 120-cell to the computational basis, and then give a simple prescription that will allow the reader to write down all the bases from the few we actually list.\\

Let rays 1-4 of the 120-cell be represented by the vectors $1=(1,0,0,0), 2=(0,1,0,0), 3=(0,0,1,0)$ and $4=(0,0,0,1)$. These rays are mutually orthogonal and form  a basis (the ``computational" basis) that we will denote 1 2 3 4. Let $U,V$ and $W$ be the orthogonal matrices

\begin{equation}
U =
\left( \begin{array}{cccc}
\frac{1}{2} & \frac{1}{2} & \frac{1}{2} & -\frac{1}{2} \\
\frac{1}{2} & \frac{1}{2} & -\frac{1}{2} & \frac{1}{2} \\
\frac{1}{2} & -\frac{1}{2} & \frac{1}{2} & \frac{1}{2} \\
\frac{1}{2} & -\frac{1}{2} & -\frac{1}{2} & -\frac{1}{2} \end{array} \right)
\end{equation}

\begin{equation}
V =
\left( \begin{array}{cccc}
\frac{\tau}{2} & 0 & -\frac{1}{2} & \frac{1}{2\tau}\\
0 & \frac{\tau}{2} & -\frac{1}{2\tau} & -\frac{1}{2} \\
\frac{1}{2} & \frac{1}{2\tau} & \frac{\tau}{2} & 0 \\
-\frac{1}{2\tau} & \frac{1}{2} & 0 & \frac{\tau}{2} \end{array} \right)
\end{equation}

\begin{equation}
W =
\left( \begin{array}{cccc}
\frac{1}{2\tau} & -\frac{\tau}{2} & 0 & \frac{1}{2}\\
\frac{\tau}{2} & \frac{1}{2\tau} & \frac{1}{2} & 0 \\
0 & -\frac{1}{2} & \frac{1}{2\tau} & -\frac{\tau}{2} \\
-\frac{1}{2} & 0 & \frac{\tau}{2} & \frac{1}{2\tau} \end{array} \right),
\end{equation} \

\noindent
where $\tau = (1+\surd 5)/2$ is the golden ratio. These matrices represent four-dimensional rotations of period 3, 5 and 5, respectively, so that $U^{3}=V^{5}=W^{5}=I$, where $I$ is the $4\times 4$ identity matrix\footnote{Since $V$ and $W$ are symmetry operations of the 120-cell, they can be described by the permutations they perform on its vertices: $V$ replaces the ray $i$ by the ray $(i+60)$ mod 300, while $W$ replaces ray $i$ by $i+12$ if $60n < i \leq 60n+48$ for $n=0,1,2,3,4$, or by $i-48$ otherwise. The operator $U$ also performs a permutation, but it cannot be described simply.}. The other 296 rays of the 120-cell can be obtained by applying products of powers of $U$, $V$ and $W$ to rays 1-4 in the manner described by the equation

\begin{equation}
|60n+12m+4l+i\rangle = W^{n}V^{m}U^{l}|i\rangle
\end{equation}

\noindent
where $i=1,2,3,4$, $l=0,1,2$ and $m,n=0,1,2,3,4$, and $|j\rangle$ (with $j = 1,\cdots,300$) is ray $j$ expressed as a four-component column vector.\\

\begin{table}[htp]
\addtolength{\tabcolsep}{-2.3pt}
\begin{center}
\begin{tabular}{|c|cccc|cccc|cccc|cccc|cccc|}
\hline
&\multicolumn{4}{|c|}{A}&\multicolumn{4}{|c|}{B}&\multicolumn{4}{|c|}{C}&
\multicolumn{4}{|c|}{D}&\multicolumn{4}{|c|}{E}\\
\hline
\multirow{3}{*}{A\'}&1&2&3&4&61&62&63&64&121&122&123&124&181&182&183&184&241&242&243&244\\[-1.5pt]
&5&6&7&8&65&66&67&68&125&126&127&128&185&186&187&188&245&246&247&248\\[-1.5pt]
&9&10&11&12&69&70&71&72&129&130&131&132&189&190&191&192&249&250&251&252\\
\hline
\multirow{3}{*}{B\'}&13&14&15&16&73&74&75&76&133&134&135&136&193&194&195&196&253&254&255&256\\[-1.5pt]
&17&18&19&20&77&78&79&80&137&138&139&140&197&198&199&200&257&258&259&260\\[-1.5pt]
&21&22&23&24&81&82&83&84&141&142&143&144&201&202&203&204&261&262&263&264\\
\hline
\multirow{3}{*}{C\'}&25&26&27&28&85&86&87&88&145&146&147&148&205&206&207&208&265&266&267&268\\[-1.5pt]
&29&30&31&32&89&90&91&92&149&150&151&152&209&210&211&212&269&270&271&272\\[-1.5pt]
&33&34&35&36&93&94&95&96&153&154&155&156&213&214&215&216&273&274&275&276\\
\hline
\multirow{3}{*}{D\'}&37&38&39&40&97&98&99&100&157&158&159&160&217&218&219&220&277&278&279&280\\[-1.5pt]
&41&42&43&44&101&102&103&104&161&162&163&164&221&222&223&224&281&282&283&284\\[-1.5pt]
&45&46&47&48&105&106&107&108&165&166&167&168&225&226&227&228&285&286&287&288\\
\hline
\multirow{3}{*}{E\'}&49&50&51&52&109&110&111&112&169&170&171&172&229&230&231&232&289&290&291&292\\[-1.5pt]
&53&54&55&56&113&114&115&116&173&174&175&176&233&234&235&236&293&294&295&296\\[-1.5pt]
&57&58&59&60&117&118&119&120&177&178&179&180&237&238&239&240&297&298&299&300\\
\hline
\end{tabular}
\end{center}
\caption{The 300 rays of the 120-cell, grouped together in blocks of 12 rays each. Each block defines a 24-cell, with each row of four rays within a block defining a basis. Each row or column of blocks defines a 600-cell, with the 600-cells defined by the columns being labeled  $A,\cdots,E$ and those defined by the rows being labeled $A^{'},\cdots,E^{'}$. Each 24-cell in this table can be labeled by a pair of letters, one unprimed and the other primed, of the two 600-cells to which it belongs.}
\vskip-20pt
\label{tab1}
\end{table}

The buildup of the rays described by (4) can be understood as follows. The operators $U$ and $U^{2}$ act on the basis 1 2 3 4 to yield the bases 5 6 7 8 and 9 10 11 12, respectively. These three bases, shown in the top left block of Table 1, define a 24-cell whose vertices are given by the vectors $|1\rangle$-$|12\rangle$  and their inverses\footnote{The 24-cell, 600-cell and 120-cell are all centrally symmetric figures whose vertices come in antipodal pairs.}. Powers of the operator $V$ acting on this 24-cell transform it into the other 24-cells shown in the first column of Table 1. The five 24-cells in the first column of Table 1 define a 600-cell whose vertices are given by the vectors $|1\rangle$-$|60\rangle$  and their inverses. Powers of $W$ acting on this 600-cell then give the four 600-cells represented by the other columns of Table 1. Remarkably, the rows of Table 1 also represent 600-cells. Thus Table 1 illustrates the interesting geometrical fact\footnote{See Ref. \cite{Coxeter}, p.270, where it is pointed out that the 600-cells in the rows and columns of Table 1 form a pair of enantiomorphous sets.} that the vertices of the 120-cell can be partitioned into those of five disjoint 600-cells in two different ways. We label the 600-cells corresponding to the columns of Table 1 by the unprimed letters $A,\cdots,E$ and those corresponding to the rows by the primed letters $A',\cdots,E'$. Also, we label any 24-cell in Table 1 by the unprimed and primed letters of the 600-cells to which it belongs (thus, for example, the cell in the top left corner has the label $AA'$).\\

Our construction of the 300 rays has also yielded 75 of the bases formed by them, which are exhibited in Table 1. However these rays also form 600 additional bases, which we now describe.\\

Each of the 600-cells in Table 1 has 75 bases associated with it, of which only 15 are shown in Table 1 (as one of its rows or columns). In Table 2 we show all 75 bases associated with 600-cell $A$; the bases in the first column are identical to those in the first column of Table 1, but the other 60 bases are new. The blocks of Table 2 also represent 24-cells, and this table illustrates the fact that the vertices of a 600-cell can be partitioned into those of five disjoint 24-cells in ten different ways (represented by its rows and columns). The columns of Table 2 are cycled by the period-5 rotation $V$, while its rows are cycled by the period-5 rotation

\begin{equation}
X =
\left( \begin{array}{cccc}
\frac{1}{2\tau} & \frac{1}{2} & 0 & \frac{\tau}{2}\\
-\frac{1}{2} & \frac{1}{2\tau} & \frac{\tau}{2} & 0 \\
0 & -\frac{\tau}{2} & \frac{1}{2\tau} & \frac{1}{2} \\
-\frac{\tau}{2} & 0 & -\frac{1}{2} & \frac{1}{2\tau} \end{array} \right)  .
\end{equation} \

\noindent
Unlike $V$, which is a symmetry operation of the 120-cell, $X$ is a symmetry operation of the 600-cell $A$ alone. The bases associated with the 600-cells $B$, $C$, $D$ or $E$ can be obtained by adding 60, 120, 180 or 240, respectively, to the numbers in Table 2 (which is equivalent to acting on the rays of 600-cell $A$ with powers of the operator $V$).\\

The 600-cells associated with the rows of Table 1 have very similar properties. Table 3 shows the bases associated with 600-cell $A'$; the rows are cycled by the period-5 rotation $W$ and the columns by the period-5 rotation

\begin{equation}
Y =
\left( \begin{array}{cccc}
-\frac{1}{2\tau} & -\frac{\tau}{2} & 0 & -\frac{1}{2}\\
\frac{\tau}{2} & -\frac{1}{2\tau} & \frac{1}{2} & 0 \\
0 & -\frac{1}{2} & -\frac{1}{2\tau} & \frac{\tau}{2} \\
\frac{1}{2} & 0 & -\frac{\tau}{2} & -\frac{1}{2\tau} \end{array} \right)  ,
\end{equation} \

\noindent
which, similar to $X$, is a symmetry operation of this 600-cell alone, and not of the whole 120-cell. Adding 12,24,36 or 48 to the numbers in Table 3 (which is equivalent to acting on the rays of 600-cell $A'$ with powers of the operator $W$) gives the bases associated with the 600-cells $B'$,$C'$,$D'$ or $E'$, respectively.\\

In summary, the 675 bases formed by the rays of the 120-cell are obtained by adding $60n$ to the entries in Table 2 and $12n$ to the entries in Table 3 for $n=0,1,2,3$ or $4$. This actually leads to $75 \times 10 = 750$ bases, but the 75 special bases of Table 1 are each generated twice in this process (once as part of an unprimed 600-cell and once as part of a primed one), and so the total number of distinct bases is just 675.

\begin{table}[htp]
\addtolength{\tabcolsep}{-2.3pt}
\begin{center}
\begin{tabular}{|c|cccc|cccc|cccc|cccc|cccc|}
\hline
&\multicolumn{20}{|c|}{600-cell $A$}\\
\hline
\multirow{3}{*}&1&2&3&4&52&15&48&34&22&60&29&44&32&41&21&59&47&33&50&13\\[-1.5pt]
&5&6&7&8&57&42&31&23&26&18&55&37&39&54&19&28&24&30&43&58\\[-1.5pt]
&9&10&11&12&38&20&25&53&51&35&16&45&36&49&46&14&17&40&56&27\\
\hline
\multirow{3}{*}&13&14&15&16&4&27&60&46&34&12&41&56&44&53&33&11&59&45&2&25\\[-1.5pt]
&17&18&19&20&9&54&43&35&38&30&7&49&51&6&31&40&36&42&55&10\\[-1.5pt]
&21&22&23&24&50&32&37&5&3&47&28&57&48&1&58&26&29&52&8&39\\
\hline
\multirow{3}{*}&25&26&27&28&16&39&12&58&46&24&53&8&56&5&45&23&11&57&14&37\\[-1.5pt]
&29&30&31&32&21&6&55&47&50&42&19&1&3&18&43&52&48&54&7&22\\[-1.5pt]
&33&34&35&36&2&44&49&17&15&59&40&9&60&13&10&38&41&4&20&51\\
\hline
\multirow{3}{*}&37&38&39&40&28&51&24&10&58&36&5&20&8&17&57&35&23&9&26&49\\[-1.5pt]
&41&42&43&44&33&18&7&59&2&54&31&13&15&30&55&4&60&6&19&34\\[-1.5pt]
&45&46&47&48&14&56&1&29&27&11&52&21&12&25&22&50&53&16&32&3\\
\hline
\multirow{3}{*}&49&50&51&52&40&3&36&22&10&48&17&32&20&29&9&47&35&21&38&1\\[-1.5pt]
&53&54&55&56&45&30&19&11&14&6&43&25&27&42&7&16&12&18&31&46\\[-1.5pt]
&57&58&59&60&26&8&13&41&39&23&4&33&24&37&34&2&5&28&44&15\\
\hline
\end{tabular}
\end{center}
\caption{The 600-cell $A$. Each row or column of blocks shows its decomposition into five disjoint 24-cells, with the first column being identical to that in Table \ref{tab1}. There are three bases in each 24-cell, and therefore 75 bases in all. The rows of blocks are cycled by the period-5 operation $W$ of Eq.(3), which simply has the effect of adding 12 to any ray number, modulo 60. The columns are cycled by the period-5 operation $X$ of Eq.(5) (whose permutation of the 60 rays is easily picked out). Adding 60, 120, 180 or 240 to the numbers in this table gives the basis tables of the 600-cells $B,C,D$ or $E$, respectively.}
\vskip-20pt
\label{tab2}
\end{table}

\begin{table}[htp]
\addtolength{\tabcolsep}{-2.3pt}
\begin{center}
\begin{tabular}{|c|cccc|cccc|cccc|cccc|cccc|}
\hline
&\multicolumn{20}{|c|}{600-cell $A'$}\\
\hline
\multirow{3}{*}&1&2&3&4&61&62&63&64&121&122&123&124&181&182&183&184&241&242&243&244\\[-1.5pt]
&5&6&7&8&65&66&67&68&125&126&127&128&185&186&187&188&245&246&247&248\\[-1.5pt]
&9&10&11&12&69&70&71&72&129&130&131&132&189&190&191&192&249&250&251&252\\
\hline
\multirow{3}{*}&127&242&186&64&187&2&246&124&247&62&6&184&7&122&66&244&67&182&126&4\\[-1.5pt]
&121&72&251&183&181&132&11&243&241&192&71&3&1&252&131&63&61&12&191&123\\[-1.5pt]
&245&68&189&130&5&128&249&190&65&188&9&250&125&248&69&10&185&8&129&70\\
\hline
\multirow{3}{*}&71&182&252&124&131&242&12&184&191&2&72&244&251&62&132&4&11&122&192&64\\[-1.5pt]
&247&190&129&66&7&250&189&126&67&10&249&186&127&70&9&246&187&130&69&6\\[-1.5pt]
&61&243&125&188&121&3&185&248&181&63&245&8&241&123&5&68&1&183&65&128\\
\hline
\multirow{3}{*}&249&122&70&184&9&182&130&244&69&242&190&4&129&2&250&64&189&62&10&124\\[-1.5pt]
&191&248&65&132&251&8&125&192&11&68&185&252&71&128&245&12&131&188&5&72\\[-1.5pt]
&187&126&241&63&247&186&1&123&7&246&61&183&67&6&121&243&127&66&181&3\\
\hline
\multirow{3}{*}&185&62&128&244&245&122&188&4&5&182&248&64&65&242&8&124&125&2&68&184\\[-1.5pt]
&69&123&181&250&129&183&241&10&189&243&1&70&249&3&61&130&9&63&121&190\\[-1.5pt]
&131&192&67&246&191&252&127&6&251&12&187&66&11&72&247&126&71&132&7&186\\
\hline
\end{tabular}
\end{center}
\caption{The 600-cell $A'$. Each row or column of blocks shows its decomposition into five disjoint 24-cells, with the first row being identical to that in Table \ref{tab1}. There are three bases in each 24-cell, and therefore 75 bases in all. The columns are cycled by the period-5 permutation $V$ of Eq.(2), which has the effect of adding 60 to any ray number, modulo 300. The rows are cycled by the period-5 operation $Y$ of Eq.(6) (whose permutation of the 60 rays in this table is easily picked out). Adding 12, 24, 36 or 48 to the numbers in this table generates the basis tables of the 600-cells $B',C',D'$ or $E'$, respectively.}
\vskip-20pt
\label{tab3}
\end{table}

\section{\label{sec:3} Parity proofs in the 120-cell}

Any subset of the 675 bases of the 120-cell provides a ``parity proof" of the KS theorem if (a) the number of bases in the subset is odd, and (b) each ray occurring in these bases occurs in an even number of them. Such a set of rays and bases provides a proof of the KS theorem because it is impossible to assign noncontextual \footnote{A noncontextual value assignment to a ray is one in which the ray is assigned the same value in all the bases in which it occurs.} values of 0 or 1 to each of the rays in such a way that each basis has exactly one ray assigned the value 1 in it. The term ``parity proof" is used because of the odd-even conflict in conditions (a) and (b) used to establish the theorem.\\

The task of finding parity proofs in the 120-cell thus reduces to that of identifying subsets of its bases satisfying conditions (a) and (b). We have developed a computer program that does this for any set of rays and bases given a target number of bases in the parity proof. The program begins from a ``seed" basis (which can be chosen at will) and adds on further bases in a calculated manner until the target number of bases is reached and conditions (a) and (b) are either satisfied, in which case one gets a parity proof, or not satisfied, in which case the search turns up empty. If a proof is found, the program checks it to see if it is critical, i.e., whether it fails if even a single basis is dropped. Searches are made for all basis sizes starting at the low end and going up. However the search can become prohibitively slow for large numbers of bases or when rays of large multiplicities are involved.\\

The 675 bases of the 120-cell constitute too large a search space for our program to operate efficiently in. We therefore had to find ways of whittling down these bases to smaller subsets that would be large enough to contain a significant store of parity proofs and at the same time small enough to be searched quickly. We found that we could generate such subsets simply by picking a certain number of 24-cells in Table 1 and dropping all the bases containing any of the rays in these 24-cells. Table 4 lists several subsets of the 675 bases generated by this procedure, with the first column indicating the 24-cells whose rays are dropped and the second column the symbols of the resulting ray-basis sets. The important point about these reduced sets is that they all span more than one 600-cell, making it possible for them to contain parity proofs that also span more than one 600-cell. We found that all the reduced sets in Table 4 do yield proofs of this kind. As one example, Table 5 lists the 102 different types of parity proofs contained within the last reduced set of Table 4.\\

\begin{table}[ht]
\centering 
\begin{tabular}{|c | c |} 
\hline 
 Canceled 24-cells & Remaining Rays-Bases  \\
\hline
$A'A,A'B,A'C,A'D,A'E$ & $240_{6}-360_4$\\
\hline
$A'A,A'B,A'C,A'D,B'E$ & $12_{2}48_{3}180_{6}-312_4$\\
\hline
$A'A,A'B,A'C,B'D,B'E$ & $60_{2}180_{6}-300_4$\\
\hline
$A'A,A'B,A'C,B'D,C'E$ & $24_{2}96_{3}120_{6}-264_4$\\
\hline
$A'A,A'B,B'C,B'D,C'E$ & $72_{2}48_{3}120_{6}-252_4$\\
\hline
$A'A,A'B,B'C,C'D,D'E$ & $36_{2}144_{3}60_{6}-216_4$\\
\hline
$A'A,A'B,A'C,A'D,A'E$ & \\
$B'B,B'C,B'D,B'E$ & \\
$C'B,C'C,C'D,C'E$ &$36_{2}48_{5}12_{6}-96_4$\\
$D'B,D'C,D'D,D'E$ & \\
\hline
\end{tabular}
\caption{If all the rays belonging to the 24-cells in the first column are dropped, along with all the bases in which one or more of these rays occur, the remaining rays and bases form the reduced sets indicated in the second column. Each of these reduced sets contains a large number of parity proofs that we have found using a computer program.}
\label{tab4} 
\end{table}

\begin{table}[ht]
\centering 
\begin{tabular}{|c | c |} 
\hline 
 Number of bases & Parity proofs  \\
\hline
$19$ & $38_{2}$\\
\hline
$21$ & $42_{2}$\\
\hline
$23$ & $46_{2},44_{2}1_{4},42_{2}2_{4}$\\
\hline
$25$ & $50_{2},48_{2}1_{4},46_{2}2_{4}$\\
\hline
$27$ & $54_{2},52_{2}1_{4},50_{2}2_{4},48_{2}3_{4},46_{2}4_{4}$\\
\hline
$29$ & $58_{2},\cdots,46_{2}6_{4}$\\
\hline
$31$ & $62_{2},\cdots,46_{2}8_{4}$\\
\hline
$33$ & $66_{2},\cdots,46_{2}10_{4}$\\
\hline
$35$ & $70_{2},\cdots,46_{2}12_{4}$\\
\hline
$37$ & $74_{2},\cdots,46_{2}14_{4}$\\
\hline
$39$ & $78_{2},\cdots,48_{2}15_{4}$\\
\hline
$41$ & $82_{2},\cdots,48_{2}17_{4}$\\
\hline
\end{tabular}
\caption{Parity proofs contained within the last reduced set of Table 4. For the number of bases shown in the first column, the second column shows the ray signatures (with multiplicities as subscripts) of the various parity proofs that exist. As the number of bases increases, the proofs can come in a variety of types, beginning with only rays of multiplicity 2 and progressing to proofs with a steadily increasing number of rays of multiplicity 4 (the dots $\cdots$ indicate a range of proofs in which two rays of multiplicity 2 are traded for one ray of multiplicity 4 as one proceeds from left to right). There are 102 different proofs in this table, all of which are critical and span more than one 600-cell.}
\label{table5} 
\end{table}

\begin{table}[ht]
\centering 
\begin{tabular}{| c | c | c | c | c | c | c | c |} 
\hline 
$AB'$ & $13$ $14$ $15$ $16$ & $AC'$ & $33$ $34$ $35$ $36$ & $AD'$ & $45$ $46$ $47$ $48$ & $BE'$ & $109$ $110$ $111$ $112$ \\
\hline
$DE'$ & $233$ $234$ $235$ $236$ & $A$ & $52$ $15$ $48$ $34$ & $A$ & $51$ $35$ $16$ $45$ & $A$ & $47$ $33$ $50$ $13$ \\
\hline
$A$ & $36$ $49$ $46$ $14$ & $E'$ & $49$ $300$ $179$ $111$ & $E'$ & $235$ $50$ $294$ $172$ & $E'$ & $169$ $120$ $299$ $231$ \\
\hline
$E'$ & $169$ $51$ $233$ $296$ & $E'$ & $299$ $110$ $180$ $52$ & $E'$ & $119$ $230$ $300$ $172$ & $E'$ & $53$ $230$ $296$ $112$ \\
\hline
$E'$ & $119$ $180$ $55$ $234$ & $E'$ & $179$ $236$ $53$ $120$ & $E'$ & $55$ $294$ $109$ $231$ & $ $ & $ $ $ $ $ $ $ $ \\
\hline
\end{tabular}
\caption{A $38_{2}$-$19_{4}$ parity proof, involving 38 rays that each occur twice among 19 bases. The 600-cell to which any basis belongs is indicated to its left, with a pair of letters being used for the special bases that belong to a pair of 600-cells.}
\label{table6} 
\end{table}

\begin{table}[ht]
\centering 
\begin{tabular}{| c | c | c | c | c | c | c | c |} 
\hline 
$AB'$ & $13$ $14$ $15$ $16$ & $AC'$ & $33$ $34$ $35$ $36$ & $AD'$ & $45$ $46$ $47$ $48$ & $AE'$ & $57$ $58$ $59$ $60$ \\
\hline
$BE'$ & $109$ $110$ $111$ $112$ & $DE'$ & $233$ $234$ $235$ $236$ & $A$ & $52$ $15$ $48$ $34$ & $A$ & $51$ $35$ $16$ $45$ \\
\hline
$A$ & $47$ $33$ $50$ $13$ & $A$ & $36$ $49$ $46$ $14$ & $E'$ & $49$ $300$ $179$ $111$ & $E'$ & $235$ $50$ $294$ $172$ \\
\hline
$E'$ & $289$ $240$ $119$ $51$ & $E'$ & $113$ $236$ $57$ $298$ & $E'$ & $115$ $230$ $174$ $52$ & $E'$ & $109$ $60$ $239$ $171$ \\
\hline
$E'$ & $49$ $231$ $113$ $176$ & $E'$ & $239$ $50$ $120$ $292$ & $E'$ & $119$ $230$ $300$ $172$ & $E'$ & $59$ $120$ $295$ $174$ \\
\hline
$E'$ & $177$ $231$ $289$ $58$ & $E'$ & $233$ $110$ $176$ $292$ & $E'$ & $179$ $240$ $115$ $294$ & $E'$ & $295$ $234$ $49$ $171$ \\
\hline
$E'$ & $171$ $50$ $298$ $112$ & $ $ & $ $ $ $ $ $ $ $ & $ $ & $ $ $ $ $ $ $ $ & $ $ & $ $ $ $ $ $ $ $ \\
\hline
\end{tabular}
\caption{A $46_{2}2_{4}$-$25_{4}$ parity proof. Rays $49$ and $50$ occur four times among the bases, and all the other rays occur twice each. The label(s) of the 600-cell(s) to which each basis belongs is indicated to its left.}
\label{table7} 
\end{table}

\begin{table}[ht]
\centering 
\begin{tabular}{| c | c | c | c | c | c | c | c |} 
\hline 
$AB'$ & 13 14 15 16 & $AC'$ & 33 34 35 36 & $AD'$ & 41 42 43 44 & $AD'$ & 45 46 47 48  \\
\hline
$BB'$ & 73 74 75 76 & $BC'$ & 93 94 95 96 & $BE'$ & 109 110 111 112 & $CA'$ & 121 122 123 124 \\
\hline
$CC'$ & 145 146 147 148 & $DD'$ & 221 222 223 224 & $EE'$ & 293 294 295 296 & $A$ & 52 15 48 34 \\
\hline
$A$ &51 35 16 45 & $A$ &47 33 50 13 & $A$ &36 49 46 14 & $B$ &112 75 108 94  \\
\hline
$B$ &111 95 76 105 & $B$ &107 93 110 73 & $B$ &96 109 106 74 & $C$ &124 147 180 166  \\
\hline
$C$ &123 167 148 177 & $C$ &179 165 122 145 & $C$ &168 121 178 146 & $D'$ &221 44 165 106  \\
\hline
$D'$ &223 166 105 42 & $D'$ &107 168 43 222 & $D'$ &167 224 41 108 & $E'$ &49 300 179 111  \\
\hline
$E'$ &173 296 117 58 & $E'$ &235 50 294 172 & $E'$ &229 180 59 291 & $E'$ &289 240 119 51  \\
\hline
$E'$ &115 230 174 52 & $E'$ &169 120 299 231 & $E'$ &235 178 117 54 & $E'$ &229 111 293 56  \\
\hline
$E'$ &119 230 300 172 & $E'$ &59 120 295 174 & $E'$ &177 231 289 58 & $E'$ &299 56 173 240  \\
\hline
$E'$ &115 54 169 291 &  & &  & &  &  \\
\hline

\end{tabular}
\caption{A $80_{2}1_{4}$-$41_{4}$ parity proof. Ray $111$ occurs four times among the bases, and all the other rays occur twice each. The label(s) of the 600-cell(s) to which each basis belongs is indicated to its left.}
\label{table8} 
\end{table}

Tables 6-8 show three explicit examples of the parity proofs listed in Table 5. Each proof spans a number of distinct 600-cells and each is also critical, as the reader may verify. \\

We have not estimated how many different types of parity proofs there are in the 120-cell (that do not lie entirely within a single 600-cell). We know there are no proofs of less than 19 bases and we have not found any with more than 41 bases, but we cannot be sure about the upper limit because our searches have been limited to only the reduced sets in Table 4. However two facts might be mentioned: the first is that a given ray-basis symbol often has a number of distinct (i.e. unitarily inequivalent) proofs associated with it, and the second is that each proof generally has hundreds or thousands of replicas under symmetry. Taking both these facts into account, we estimate that there are probably over a million genuinely new parity proofs in the 120-cell that are not contained in any of the smaller polytopes in it.\\

\section{\label{sec:4} Discussion}

This paper has used the symmetries of the 120-cell to give a simple construction of the 300 rays and 675 bases associated with it (see Tables 1-3); it has identified several subsets of the 675 bases that are quickly searched for parity proofs (see Table 4); it has given a detailed account of the parity proofs in one of the subsets (see Table 5); and it has listed three explicit examples of the parity proofs (see Tables 6-8) so that any reader can see that they work as advertised. The framework established in this paper can be used by others who wish to view carry out a more exhaustive search for parity proofs in the 120-cell.\\

As mentioned in the introduction, parity proofs are interesting because they can be used to devise experimental tests of quantum contextuality and also have a variety of applications in quantum information processing. The 120-cell is the most complicated member of a family that includes the 600-cell and the 24 Peres rays, but it abounds in many parity proofs that are distinctly its own and not contained in any of the smaller polytopes. The 120-cell (like the 600-cell and the Peres rays) can be realized experimentally using a pair of qubits. From Eq(4) it is clear that all the rays and bases of the 120-cell can be built up from the computational basis if one has the ability to implement the gates represented by the operators $U,V,W,X,Y$ and their powers. This is a considerable experimental challenge, but it is not beyond the realm of possibility. It would be nice to find other examples of tasks that can be accomplished within the finite (but fairly large) universe of states and bases provided by the 120-cell, as that might further spur its experimental realization. Whether there are any practical applications or not, the proofs of quantum contextuality made possible by the four-dimensional regular polytopes represent a charming encounter between classical geometry and quantum physics that does credit to both.\\


\end{document}